# Visualising how a fuse works

M. M. J. French

Physics Department, Clifton College, Guthrie Road, Clifton, Bristol

E-mail: mail@matthewfrench.net

In this brief article I describe an experiment to illustrate how a fuse works. I have used this as part of lessons for my year 11 classes to demonstrate how an electrical fuse 'blows' when too high a current passes through it.

A Variac (also known as an autotransformer) is connected to the mains and to the primary coil (approximately 100 turns) of a transformer with a laminated iron core. The secondary coil of the transformer (approximately 8 turns) is connected to two thick metal terminals with screw attachments. A thick piece of wire (approximately 1-1.5mm diameter – actually a piece of wire coat hanger) is connected between the two screw attachments across the secondary coil. See figure 1. When the mains supply to the Variac is turned on the transformer acts to step down the voltage and step up the current. As the Variac output voltage (i.e. the voltage across the primary coil of the transformer) is turned up to around 240V, the current flowing through the wire (connected across the secondary coil of the transformer) increases so much that it began to heat up (since power = current² x resistance). The wire will soon begin to glow, first red and then white before it melts – exactly what happens with a fuse. See figure 2. This provides a great way for pupils to visualise what happens when a fuse blows. It can also be used to demonstrate the transformer effect: stepping down a voltage and so stepping up the current if power is conserved (if the transformer is 100% efficient).

The safety aspects of this demonstration need to be considered carefully by the individual teacher before it is attempted. There are obvious dangers due to the high voltages used and proper protected high voltage leads should be used for all the connections. The circuit should not be dismantled unless unplugged from the mains. The wire on the secondary coil (and in the attached circuit) needs to be very thick as it must not melt when the coat hanger wire does so with the same current flowing through it. The equipment should be placed on heat proof mats to protect the desk and safety screens should be used to protect both the pupils and the teacher. The wire which melts will remain very hot for some considerable time after the demonstration is complete and pupils should not be allowed to touch it even as they leave at the end of the lesson.

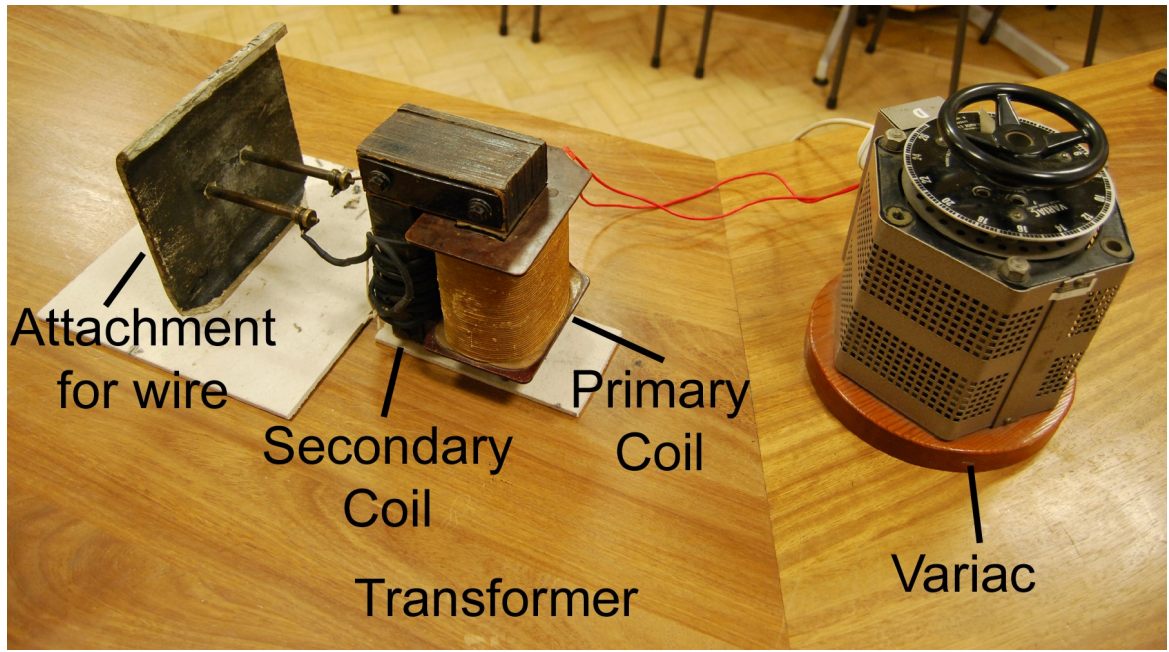

Figure 1: Diagram of the equipment which shows the Variac (right) which is connected to the transformer (middle) and the coat hanger wire (left). The coat hanger wire is attached through a shield made from a heat proof mat and is placed on a heat proof mat. Safety screens should be in place when this equipment is used to protect both the pupils and the teacher.

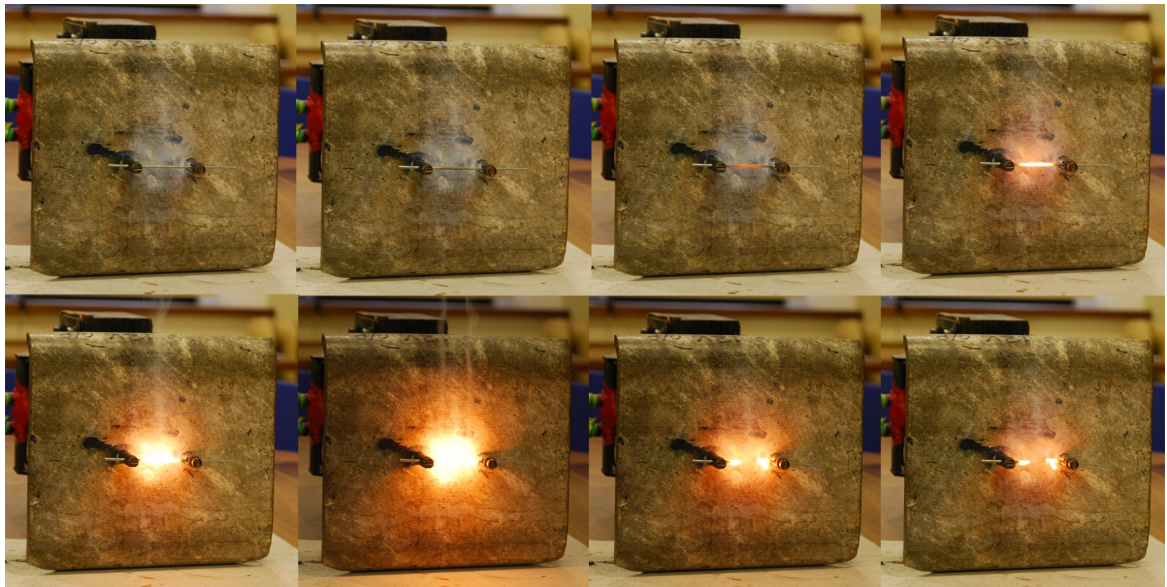

Figure 2: Composite image showing 8 frames in succession of the thick coat hanger wire as it melts due to the heat generated by the high current passing through it.